# Power System Fault Diagnosis with Quantum Computing and Efficient Gate Decomposition

Xiang Fei[1], Huan Zhao[2], Xiyuan Zhou[3], Junhua Zhao[3], Ting Shu[4], Fushuan Wen[5]

[1]School of Data Science, The Chinese University of Hong Kong (Shenzhen), Shenzhen 518172, China
[2]School of Electrical and Electronic Engineering, Nanyang Technological University, Singapore 639798, Singapore
[3]School of Science and Engineering, The Chinese University of Hong Kong (Shenzhen), Shenzhen 518172, China
[4]Shenzhen Institute of Meteorological Innovation, Shenzhen 518038, China
[5]College of Electrical Engineering, Zhejiang University, Hangzhou 310027, China

*Abstract*—**Power system fault diagnosis is crucial for identifying the location and causes of faults and providing decision-making support for power dispatchers. However, most classical methods suffer from significant time-consuming, memory overhead, and computational complexity issues as the scale of the power system concerned increases. With rapid development of quantum computing technology, the combinatorial optimization method based on quantum computing has shown certain advantages in computational time over existing methods. Given this background, this paper proposes a quantum computing based power system fault diagnosis method with the Quantum Approximate Optimization Algorithm (QAOA). The proposed method reformulates the fault diagnosis problem as a Hamiltonian by using Ising model, which completely preserves the coupling relationship between faulty components and various operations of protective relays and circuit breakers. Additionally, to enhance problem-solving efficiency under current equipment limitations, the symmetric equivalent decomposition method of multi-z-rotation gate is proposed. Furthermore, the small probability characteristics of power system events is utilized to reduce the number of qubits. Simulation results based on the test system show that the proposed methods can achieve the same optimal results with a faster speed compared with the classical higher-order solver provided by D-Wave.**

Power system faults could result in widespread power outages, leading to significant economic losses and safety concerns [1]. To identify faulty equipment and process, the power system fault diagnosis analyzes the action behaviors of protective relays (PRs) and circuit breakers (CBs) based on alarm information collected by the control center. The fault diagnosis result helps to reconstruct the fault process and provides decision-making support for dispatchers. However, with the increasing complexity of modern power grids, power system faults are displaying more intricate patterns [2]. Hence, there is an urgent need for faster and more reliable power system fault diagnosis technologies to support power system operation.

Various kinds of methods have been developed for power system fault diagnosis and can be generally divided into two main categories, data-driven and optimization-based methods. For the data-driven methods, expert systems (ESs) and artificial neural networks (ANNs) are most widely used. The ES captures the behavior of the protection system through powerful logical reasoning [3], [4]. However, the learning capability and fault tolerance of ES-based methods are weak, and the establishment and maintenance of the knowledge base are challenging. The ANN-based methods utilize sample data to model the complex internal logic of the protection system and are of advantages in fast diagnosis speed, strong learning ability, and good fault tolerance. Various types of ANNS and related algorithms, such as the radial basis function (RBF) [5], multilayer perception (MLP) [6], support vector machine (SVM) [7], and parameter selection [8], have been applied for automatic fault detection and diagnosis. However, the ANN-based methods have limitations due to the dependency on the fault sample volume and quality. Also, it is difficult to establish a complete set of fault samples due to the small probability of power system faults.

The optimization-based method models the power system fault diagnosis as a binary integer programming problem and uses optimization technology to solve it, which has a strict mathematical foundation and theoretical basis. Power system fault diagnosis can be expressed as an NP-hard problem [9], [10] and various kinds of heuristic algorithms are often used to obtain approximate optimal solutions, such as tabu search (TS) [11], particle swarm optimization algorithm (PSO) [12], generic algorithm-II (NSGA-II) [13], binary coded brain storm optimization (BCBSO) [14]. Although these classical heuristic algorithms can effectively diagnose faulty components, they still suffer limitations in terms of inaccurate solution, time-consuming, memory overhead, and computational complexity, especially for large-scale power grids. Meanwhile, with the continuous development of quantum computing, its computational advantages are expected to well address these challenges.

Quantum computing is considered to be a new computing paradigm with a disruptive impact on the future [15]. For combinatorial optimization problems, methods such as the quantum annealing algorithm (QA) and quantum approximate optimization algorithm (QAOA) have appeared in the field of quantum computing [16]. QA uses the quantum tunneling effect to complete the optimization process through a guided quantum evolution. QAOA is a quantum-classical hybrid algorithm based on the gate structure and has a potential exponential speedup in solving combinatorial optimization problems [17]. Although in the short term, the capabilities of quantum devices are limited by the number of qubits, gate fidelity, and error correction capabilities, QAOA has great potential to solve power system diagnosis.

QAOA has been applied to various kinds of combinatorial optimization problems, such as the maximum cut problem [18]-[20], the exact coverage problem [21]-[27], and the Hamilton path problem [28]. However, there are few researches on the application of the QAOA algorithm in power systems at present, and existing researches mainly focus on the unit commitment problem [29]-[32]. The limitation of the aforementioned research works is that they only focus on the Quadratic Unconstrained Binary Optimization (QUBO) problem, and do not consider the more intricate Polynomial Unconstrained Binary Optimization (PUBO) problem. This is largely due to the current hardware's limitations and error-proneness in implementing multi-qubit gates. However, the power system fault diagnosis problem is normally formulated as a PUBO problem [33] and this issue should be overcome. Less research has been conducted on using QAOA to solve the PUBO problem. Herrman *et al* [34] uses a global variable substitution method to convert the PUBO problem into a QUBO problem by adding auxiliary qubits, and then solves the Boolean satisfiability problem through QAOA. However, the algorithm requires a considerable number of qubits and results in a complex equivalent circuit.

Aiming at the above problems, this paper proposes a quantum computing based power system fault diagnosis method using QAOA and gate decomposition method. Four major contributions of this paper are summarized as follows. First, a quantum computing-based power system fault diagnosis framework is proposed. To the best of our knowledge, we are among the first to apply quantum computing to the power system fault diagnosis problem. Compared with the classical higher-order solver provided by D-Wave, the proposed method can obtain the same optimal results with shorter computational time, especially for large-scale power systems. Second, the Ising model considering PR and CB failed/mal operation and contradictory logic is proposed to express the Hamiltonian of power system fault diagnosis, and the quantum logic gate corresponding to the Hamiltonian is derived. Third, the Symmetric Equivalent Decomposition (SED) of the multi-z-rotation gate is proposed to efficiently solve the QAOA-based PUBO problem. Compared with the global variable substitution method using auxiliary qubits, the proposed method requires less qubits, and shows better performance with lower complexity. Fourth, a quantum circuit simplifying method is proposed by using the small probability event characteristic (SPEC) of power system faults, which reduces the high-order terms in the Hamiltonian. Simulation results shows that the proposed method can improve both operating speed and accuracy.

## Results
**Fault hypothesis**. The fault hypothesis (FH) describes the fault assumptions in a fault-tolerant system, and in this paper, we consider suspicious faulty component and the related PR and CB as hypothesis conditions [35]. Suppose that there are $n_d$ suspected faulty components in the faulty region, and the number of related PR and CB are $n_r$ and $n_c$, respectively. Then, the FH can be expressed as follows:

$$H = (D, R, C, F, M) \qquad (1)$$

where,
- $D = (d_1, \dots, d_i, \dots, d_{n_d})$ is the set of suspected faulty components. $d_i = 1$ means that the $i$th component is fault, while $d_i = 0$ means normal.
- $R = (r_1, \dots, r_i, \dots, r_{n_r})$ is the set of PRs. $r_i = 1$ means that the $i$th PR operates, while $r_i = 0$ means not operates.
- $C = (c_1, \dots, c_i, \dots, c_{n_c})$ is the set of CBs. $c_i = 1$ means that the $i$th CB is tripped, while $c_i = 0$ means not tripped.
- $F = (f_{r_1}, \dots, f_{r_{n_r}}, f_{c_1} \dots, f_{c_{n_c}})$ is the set of failed operation information. If $r_i$ or $c_i$ has performed a failed operation, then $f_{r_i}$ or $f_{c_i}$ equals to 1. Otherwise, it is 0.
- $M = (m_{r_1}, \dots, m_{r_{n_r}}, m_{c_1} \dots, m_{c_{n_c}})$ is the set of mal operation information. If $r_i$ or $c_i$ has performed a mal operation, then $m_{r_i}$ or $m_{c_i}$ equals to 1. Otherwise, it is 0.

**Fault diagnosis problem formulation**. The fault diagnosis problem can be formulated as a combinatorial optimization problem. The objective function describes the criteria and factors for judging the expected fault diagnosis [36], [37]. Based on [12], this paper completely preserves the coupling relationship between the faulty components and the operation of PR and CB. Considering the failed/mal operation of PR and CB, as well as the contradictory logic, the objective function is formulated as:

$$\min E(H) = \omega_1 \sum_{i=1}^{n_r}(r_i \oplus r_i^*) + \omega_1 \sum_{i=1}^{n_c}(c_i \oplus c_i^*) + \omega_2 \sum_{i=1}^{n_r}(r_i \oplus r_i') + \omega_2 \sum_{i=1}^{n_c}(c_i \oplus c_i') + \omega_3 \sum_{i=1}^{n_r}(f_{r_i} + m_{r_i}) + \omega_3 \sum_{i=1}^{n_c}(f_{c_i} + m_{c_i}) + \omega_4 \sum_{i=1}^{n_r}(f_{r_i} m_{r_i} + r_i f_{r_i} + \overline{r_i} m_{r_i} + p_{r_i} m_{r_i} + \overline{p_{r_i}} f_{r_i}) + \omega_4 \sum_{i=1}^{n_c}(f_{c_i} m_{c_i} + c_i f_{c_i} + \overline{c_i} m_{c_i} + p_{c_i} m_{c_i} + \overline{p_{c_i}} f_{c_i}) \qquad (2)$$

where $r_i, r_i^*, r_i', p_{r_i}$ represent the actual state, expected state, alarm, and action expectation of PR, $c_i, c_i^*, c_i', p_{c_i}$ represent the actual state, expected state, alarm, and action expectation of CB. $\oplus$ is exclusive-or operator. $\overline{\ }$ is NOT operator. $\omega_1, \omega_2, \omega_3, \omega_4$ represent the weight of each term, which must satisfy $\omega_4 \gg \omega_1, \omega_2, \omega_3$, since the logic constraint is a strong constraint condition.

The first two terms $\sum_{i=1}^{n_r}(r_i \oplus r_i^*)$ and $\sum_{i=1}^{n_c}(c_i \oplus c_i^*)$ describe the difference between the actual state and the expected state of PR and CB, respectively. The third and fourth terms $\sum_{i=1}^{n_r}(r_i \oplus r_i')$ and $\sum_{i=1}^{n_c}(c_i \oplus c_i')$ describe the difference between the actual state and the alarm state of PR and CB, respectively. The terms $\sum_{i=1}^{n_r}(f_{r_i} + m_{r_i})$ and $\sum_{i=1}^{n_c}(f_{c_i} + m_{c_i})$ describe the failed/mal operation of PR and CB, respectively. The last two terms $\sum_{i=1}^{n_r}(f_{r_i} m_{r_i} + r_i f_{r_i} + \overline{r_i} m_{r_i} + p_{r_i} m_{r_i} + \overline{p_{r_i}} f_{r_i})$ and $\sum_{i=1}^{n_c}(f_{c_i} m_{c_i} + c_i f_{c_i} + \overline{c_i} m_{c_i} + p_{c_i} m_{c_i} + \overline{p_{c_i}} f_{c_i})$ describe the number of contradictions that do not satisfy the action logic of PR and CB. The contradictory logic of the last two terms includes: failed operation with mal operation, operate with fail to operate, no operation with mal operation, incentive (expected to operate) with mal operation, no incentive with failed operation.

**Action logic model**. The action logic model is used to describe the expected state according to the protection principle. The action logic for PR and CB in this paper is shown as follows.
1) Action Logic of Main PR

Let $r_i$ be the main PR of the component $d_n$. If $d_n$ is failed ($d_n = 1$) and $r_i$ does not fail to operate, or there is a mal operation on $r_i$, then $r_i$ should operate. The action expectation and expected state of main PR are:

$$p_{r_i} = d_n \qquad (3)$$
$$r_i^* = d_n \overline{f_{r_i}} \vee m_{r_i} \qquad (4)$$

2) Action Logic of First Back-up PR

Let $r_i$ be the first back-up PR of the component $d_n$, if $d_n$ is failed and its main PR $r_j$ does not operate, then $r_i$ should operate. Consider the failed/mal operation, the action expectation and expected state of first back-up PR are:

$$p_{r_i} = d_n \overline{r_j} \qquad (5)$$
$$r_i^* = d_n \overline{r_j} \overline{f_{r_i}} \vee m_{r_i} \qquad (6)$$

3) Action Logic of Second Back-up PR

Let $r_i$ be the second back-up PR of the component $d_n$, $D(r_i, d_n)$ be the set of neighboring devices of $d_n$ within the protection range of $r_i$, $C(r_i, d_m)$ be the CBs on the path from $d_m$ to $d_n$. If the component $d_m \in D(r_i, d_n)$ is failed and there is no $c_t \in C(r_i, d_m)$ trips, then $r_i$ should operate. Considering the failed/mal operation, the action expectation and expected state of second back-up PR are:

$$p_{r_i} = \vee_{d_m \in D(r_i, d_n)}(d_m \wedge_{c_t \in C(r_i, d_m)} \overline{c_t}) \qquad (7)$$
$$r_i^* = (\vee_{d_m \in D(r_i, d_n)}(d_m \wedge_{c_t \in C(r_i, d_m)} \overline{c_t})) \overline{f_{r_i}} \vee m_{r_i} \qquad (8)$$

4) Action logic of CB

Let $R(c_i)$ be the set of PRs which can trigger $c_i$. If any $r_i \in R(c_i)$ operate, then $c_i$ should trip. Considering the failed/mal operation, the action expectation and expected state of CB are:

$$p_{c_i} = \vee_{r_i \in R(c_i)} r_i \qquad (9)$$
$$c_i^* = (\vee_{r_i \in R(c_i)} r_i) \overline{f_{c_i}} \vee m_{c_i} \qquad (10)$$

**QAOA-based fault diagnosis framework.** The proposed QAOA-based fault diagnosis framework is shown in Fig. 1. In our proposed framework, the coupling relationship between the suspected faulty components and the operation of PR and CB is completely preserved and the failed/mal operation of PR and CB, as well as the contradictory logic, is considered. To reduce the dimensionality of the original problem, the small probability event characteristic of power system failure is utilized, which can reduce the number of qubits to be used and simplify the quantum circuit. Additionally, the SED method of the multi-z-rotation gate is proved, which enables QAOA to solve the power system fault diagnosis problem in PUBO form.

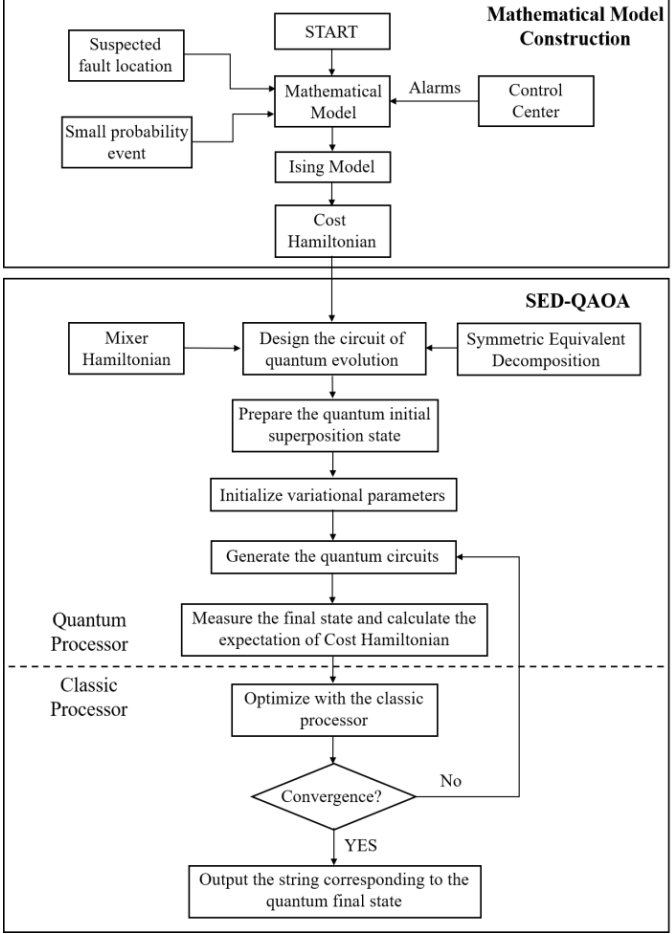

**Fig. 1 The QAOA-based fault diagnosis framework.**

The QAOA-based power system fault diagnosis framework consists of two main components: mathematical model construction and SED-QAOA. In the first part, the mathematical model of power system fault diagnosis problem is generated, and then, the model is reformulated into the form of the Ising model and cost Hamiltonian in turn. In the SED-QAOA part, the quantum circuit is created using the SED method of the multi-z-rotation gate, then the problem can be solved with QAOA. The specific procedure of each component is as follows.

In the mathematical model construction part, the power outage area is first identified to determine the suspected faulty components and related PRs and CBs, using the real-time information of the CB. The tie-line analysis method is used to identify the system topology before and after the fault, and find the difference between the two topology structures [38]. The mathematical model is then constructed, taking into account small probability event characteristics and alarms from control center. After the mathematical model is reformulated into the Ising model, the cost Hamiltonian of the fault diagnosis problem is determined.

In the SED-QAOA part, the quantum circuit is first generated with the cost Hamiltonian $H_C$, mixer Hamiltonian $H_B$, and the SED of the multi-z-rotation gate. The quantum circuit is equivalent to a parametric unitary transformation $U(H_C, H_B, \vec{\gamma}, \vec{\beta}, k)$, which is in the following form:

$$U(H_C, H_B, \vec{\gamma}, \vec{\beta}, k) = U(H_B, \beta_k)U(H_C, \gamma_k) \cdots U(H_B, \beta_1)U(H_C, \gamma_1) \quad (11)$$

where $\vec{\gamma} = (\gamma_1, \dots, \gamma_k)$ and $\vec{\beta} = (\beta_1, \dots, \beta_k)$ are parametric vectors, and $k$ is the number of alternations, also known as the level of the circuit. $U(H_C, \gamma_i)$ and $U(H_B, \beta_i)$ are two parametric unitary transformation correspond to the cost Hamiltonian and mixer Hamiltonian, which can be expressed as:

$$\begin{cases} U(H_C, \gamma_i) = e^{-i\gamma_i H_C} \\ U(H_B, \beta_i) = e^{-i\beta_i H_B} \end{cases}, i = 1, 2, \dots, k \quad (12)$$

The $H_C$ is generated from the Ising model, which is attained from the mathematical model of the problem. The derivation method is introduced in the next subsection. The mixer Hamiltonian is $\sum_{n=1}^{N} X_n$, where $N$ is the number of qubits, and $X_n$ is the Pauli-X operator acting on qubit $n$, which means that an RX gate on the $n$th qubit in the corresponding quantum circuit. The mathematical equation of the RX gate is shown as follows:

$$RX(\theta) = \begin{bmatrix} \cos\left(\frac{\theta}{2}\right) & -i\sin\left(\frac{\theta}{2}\right) \\ -i\sin\left(\frac{\theta}{2}\right) & \cos\left(\frac{\theta}{2}\right) \end{bmatrix} \quad (13)$$

After the construction of the quantum circuit, the quantum initial superposition state is prepared as $|+\rangle^{\otimes N}$. Through the quantum circuit, the quantum final state can be obtained:

$$|\psi_k\rangle = U(H_C, H_B, \vec{\gamma}, \vec{\beta}, k)|+\rangle^{\otimes N} \quad (14)$$

The quantum final state is measured by standard calculation basis and the expected value is:

$$F_k(\vec{\gamma}, \vec{\beta}) = \langle \psi_k | H_C | \psi_k \rangle \quad (15)$$

According to the above formula, the smaller the expected value $F_k(\vec{\gamma}, \vec{\beta})$ is, the higher the probability of the eigenstate corresponding to lower eigenvalue is. An eigenstate is a stable state of the quantum system that corresponds to a particular eigenvalue and refers to a possible solution in the context of QAOA. Therefore, the optimal solution of the problem can be obtained with the highest probability. The above process can be implemented with quantum processor.

And then, the expected value is minimized using classic processor:

$$(\vec{\gamma}^*, \vec{\beta}^*) = argmin_{\vec{\gamma}, \vec{\beta}} F_k(\vec{\gamma}, \vec{\beta}) \quad (16)$$

The solution to the fault diagnosis problem is obtained upon the convergence of the expectation of cost Hamiltonian. Otherwise, it is required to come back to generate a new quantum circuit that incorporates the updated parameters, and to repeat the process iteratively until convergence.

**Ising formulation of fault diagnosis problem.** The binary optimization problem can be transformed into the quantum Ising model by the spin variable quantity, and the Hamiltonian of the quantum system can be expressed by the quantum Ising model [45], [46]. In this section, the quantum Ising model of the power system fault diagnosis problem is deduced according to the mathematical model in Section II. The deduced Ising model contains higher-order terms, such as $\frac{s_r^i s_d^n s_{fr}^i s_{mr}^i}{8}$ in $r_i \oplus r_i^*$, which leads fault diagnosis to the PUBO problem. The Ising model of each term in the objective function is introduced in detail below.

Assume the index of $r_i, d_n, f_{r_i}, m_{r_i}$ in FH is $t_r^i, t_d^n, t_{fr}^i, t_{mr}^i$. By substituting the binary variable $r_i, d_n, f_{r_i}, m_{r_i} \in \{0,1\}$ with spin variable quantity $s_r^i, s_d^n, s_{fr}^i, s_{mr}^i \in \{-1,1\}$, that is:

$$r_i = \frac{1-s_r^i}{2}, d_n = \frac{1-s_d^n}{2}, f_{r_i} = \frac{1-s_{fr}^i}{2}, m_{r_i} = \frac{1-s_{mr}^i}{2} \quad (17)$$

- **The Ising model for the first term** $r_i \oplus r_i^*$
  a) If $r_i$ is the main PR of $d_n$. Then, we have: $r_i^* = d_n \overline{f_{r_i}} \vee m_{r_i}$. Therefore:
  $$r_i \oplus r_i^* = r_i(1 - d_n\overline{f_{r_i}})\overline{m_{r_i}} + \overline{r_i}(1 - (1 - d_n\overline{f_{r_i}})\overline{m_{r_i}}) \quad (18)$$
  Then, the quantum Ising model is obtained:
  $$ISING1 = \frac{1}{2} + \frac{s_r^i}{8} - \frac{3s_r^i s_{mr}^i}{8} + \frac{s_r^i s_{fr}^i}{8} - \frac{s_r^i s_d^n}{8} + \frac{s_r^i s_{fr}^i s_{mr}^i}{8} - \frac{s_r^i s_d^n s_{fr}^i}{8} - \frac{s_r^i s_d^n s_{mr}^i}{8} - \frac{s_r^i s_d^n s_{fr}^i s_{mr}^i}{8} \quad (19)$$
  b) If $r_i$ is the first back-up PR of $d_n$. Consider $r_j$ is the main PR of $d_n$. Then, we have: $r_i^* = d_n \overline{r_j} \overline{f_{r_i}} \vee m_{r_i}$. Therefore:
  $$r_i \oplus r_i^* = r_i(1 - d_n\overline{r_j}\overline{f_{r_i}})\overline{m_{r_i}} + \overline{r_i}(1 - (1 - d_n\overline{r_j}\overline{f_{r_i}})\overline{m_{r_i}}) \quad (20)$$
  Then, the quantum Ising model is obtained:
  $$ISING1 = \frac{1}{2} + \frac{s_r^i}{16} - \frac{7s_r^i s_{mr}^i}{16} + \frac{s_r^i s_{fr}^i}{16} + \frac{s_r^i s_r^j}{16} - \frac{s_r^i s_d^n}{16} + \frac{s_r^i s_{fr}^i s_{mr}^i}{16} + \frac{s_r^i s_r^j s_{fr}^i}{16} + \frac{s_r^i s_r^j s_{mr}^i}{16} - \frac{s_r^i s_d^n s_{fr}^i}{16} - \frac{s_r^i s_d^n s_{mr}^i}{16} - \frac{s_r^i s_d^n s_r^j}{16} + \frac{s_r^i s_r^j s_{fr}^i s_{mr}^i}{16} - \frac{s_r^i s_d^n s_{fr}^i s_{mr}^i}{16} - \frac{s_r^i s_d^n s_r^j s_{fr}^i}{16} - \frac{s_r^i s_d^n s_r^j s_{mr}^i}{16} - \frac{s_r^i s_d^n s_r^j s_{fr}^i s_{mr}^i}{16} \quad (21)$$
  c) If $r_i$ is the second back-up PR of $d_n$. Then, we have: $r_i^* = (\vee_{d_m \in D(r_i, d_n)}(d_m \wedge_{c_t \in C(r_i, d_m)} \overline{c_t}))\overline{f_{r_i}} \vee m_{r_i}$. Therefore:
  $$r_i \oplus r_i^* = r_i\overline{m_{r_i}}(1 - \overline{f_{r_i}}(1 - \wedge_{d_m \in D(r_i, d_n)}(1 - d_m \wedge_{c_t \in C(r_i, d_m)} \overline{c_t}))) + \overline{r_i}(1 - \overline{m_{r_i}}(1 - \overline{f_{r_i}}(1 - \wedge_{d_m \in D(r_i, d_n)}(1 - d_m \wedge_{c_t \in C(r_i, d_m)} \overline{c_t})))) \quad (22)$$
  Then, the quantum Ising model is obtained:
  $$ISING1 = \frac{1-s_r^i s_{mr}^i}{2} + \frac{s_r^i + s_r^i s_{fr}^i + s_r^i s_{mr}^i + s_r^i s_{fr}^i s_{mr}^i}{4}(1 - \prod_{d_m \in D(r_i, d_n)}(1 - \frac{1-s_d^m}{2}\prod_{c_t \in C(r_i, d_m)}\frac{1+s_c^t}{2})) \quad (23)$$

- **The Ising model for the second term** $c_j \oplus c_j^*$
  We have: $c_j^* = (\vee_{r_i \in R(c_j)} r_i)\overline{f_{c_j}} \vee m_{c_j}$. Therefore:
  $$c_j \oplus c_j^* = c_j\overline{m_{c_j}}(1 - \overline{f_{c_j}}(1 - \wedge_{r_i \in R(c_j)}\overline{r_i})) + \overline{c_j}(1 - \overline{m_{c_j}}(1 - \overline{f_{c_j}}(1 - \wedge_{r_i \in R(c_j)}\overline{r_i}))) \quad (24)$$
  Then, the quantum Ising model is obtained:
  $$ISING2 = \frac{1-s_c^j s_{mc}^j}{2} + \frac{s_c^j + s_c^j s_{fc}^j + s_c^j s_{mc}^j + s_c^j s_{fc}^j s_{mc}^j}{4}(1 - \prod_{r_i \in R(c_j)}\frac{1+s_r^i}{2}) \quad (25)$$

- **The Ising model for the third term** $r_i \oplus r_i'$
  If the alarm $r_i' = 1$, we have: $r_i \oplus r_i' = \overline{r_i}$
  Therefore, the quantum Ising model:
  $$ISING3 = \frac{1+s_r^i}{2} \quad (26)$$
  If the alarm $r_i' = 0$, then the quantum Ising model:
  $$ISING3 = \frac{1-s_r^i}{2} \quad (27)$$

- **The Ising model for the fourth term** $c_j \oplus c_j'$
  If the alarm $c_j' = 1$, we have: $c_j \oplus c_j' = \overline{c_j}$
  Therefore, the quantum Ising model:
  $$ISING4 = \frac{1+s_c^j}{2} \quad (28)$$
  If the alarm $c_j' = 0$, then the quantum Ising model:
  $$ISING4 = \frac{1-s_c^j}{2} \quad (29)$$

- **The Ising model for the fifth term** $f_{r_i} + m_{r_i}$
  The quantum Ising model:
  $$ISING5 = \frac{1-s_{fr}^i}{2} + \frac{1-s_{mr}^i}{2} \quad (30)$$

- **The Ising model for the sixth term** $f_{c_j} + m_{c_j}$
  The quantum Ising model:
  $$ISING6 = \frac{1-s_{fc}^j}{2} + \frac{1-s_{mc}^j}{2} \quad (31)$$

- **The Ising model for the seventh term** $f_{r_i}m_{r_i} + r_i f_{r_i} + \overline{r_i}m_{r_i} + p_{r_i}m_{r_i} + \overline{p_{r_i}}f_{r_i}$
  a) If $r_i$ is the main PR of $d_n$. Then, we have: $p_{r_i} = d_n$.
  Therefore:
  $$f_{r_i}m_{r_i} + r_i f_{r_i} + \overline{r_i}m_{r_i} + p_{r_i}m_{r_i} + \overline{p_{r_i}}f_{r_i} = f_{r_i}m_{r_i} + r_i f_{r_i} + \overline{r_i}m_{r_i} + d_n m_{r_i} + \overline{d_n}f_{r_i} \quad (32)$$
  Then, the quantum Ising model:
  $$ISING7 = \frac{1}{4}(5 - 3s_{fr}^i - 3s_{mr}^i + s_{fr}^i s_{mr}^i + s_r^i s_{fr}^i - s_r^i s_{mr}^i + s_d^n s_{mr}^i - s_d^n s_{fr}^i) \quad (33)$$
  b) If $r_i$ is the first back-up PR of $d_n$. Consider $r_j$ is the main PR of $d_n$. Then, we have: $p_{r_i} = d_n\overline{r_j}$. Therefore:
  $$f_{r_i}m_{r_i} + r_i f_{r_i} + \overline{r_i}m_{r_i} + p_{r_i}m_{r_i} + \overline{p_{r_i}}f_{r_i} = f_{r_i}m_{r_i} + r_i f_{r_i} + \overline{r_i}m_{r_i} + d_n\overline{r_j}m_{r_i} + (1 - d_n\overline{r_j})f_{r_i} \quad (34)$$
  Then, the quantum Ising model:
  $$ISING7 = \frac{1}{8}(10 - 7s_{fr}^i - 5s_{mr}^i + 2s_{fr}^i s_{mr}^i + 2s_r^i s_{fr}^i - 2s_r^i s_{mr}^i + s_d^n s_{mr}^i - s_r^j s_{mr}^i + s_r^j s_{fr}^i - s_d^n s_{fr}^i + s_d^n s_r^j s_{mr}^i - s_d^n s_r^j s_{fr}^i) \quad (35)$$
  c) If $r_i$ is the second back-up PR of $d_n$. Then, we have: $p_{r_i} = \vee_{d_m \in D(r_i, d_n)}(d_m \wedge_{c_t \in C(r_i, d_m)} \overline{c_t})$. Therefore:
  $$f_{r_i}m_{r_i} + r_i f_{r_i} + \overline{r_i}m_{r_i} + p_{r_i}m_{r_i} + \overline{p_{r_i}}f_{r_i} = f_{r_i}m_{r_i} + r_i f_{r_i} + \overline{r_i}m_{r_i} + m_{r_i}\vee_{d_m \in D(r_i, d_n)}(d_m \wedge_{c_t \in C(r_i, d_m)} \overline{c_t}) + f_{r_i}(1 - \vee_{d_m \in D(r_i, d_n)}(d_m \wedge_{c_t \in C(r_i, d_m)} \overline{c_t})) \quad (36)$$
  Then, the quantum Ising model:
  $$ISING7 = \frac{5}{4} - \frac{s_{fr}^i}{2} - s_{mr}^i + \frac{s_{fr}^i s_{mr}^i}{4} + \frac{s_r^i s_{fr}^i}{4} - \frac{s_r^i s_{mr}^i}{4} + \frac{s_{mr}^i - s_{fr}^i}{2}\prod_{d_m \in D(r_i, d_n)}(1 - \frac{1-s_d^m}{2}\prod_{c_t \in C(r_i, d_m)}\frac{1+s_c^t}{2}) \quad (37)$$

- **The Ising model and Hamiltonian for the eighth term** $f_{c_j}m_{c_j} + c_j f_{c_j} + \overline{c_j}m_{c_j} + p_{c_j}m_{c_j} + \overline{p_{c_j}}f_{c_j}$
  We have: $p_{c_j} = \vee_{r_i \in R(c_j)} r_i$. Therefore:

$$f_{c_j}m_{c_j} + c_jf_{c_j} + \overline{c_j}m_{c_j} + p_{c_j}m_{c_j} + \overline{p_{c_j}}f_{c_j} = f_{c_j}m_{c_j} + c_jf_{c_j} + \overline{c_j}m_{c_j} + m_{c_j}\left(1 - \wedge_{r_i \in R(c_j)}\overline{r_i}\right) + f_{c_j}\wedge_{r_i \in R(c_j)}\overline{r_i} \tag{38}$$

Then, the quantum Ising model:

$$ISING8 = \frac{1}{4}(5 - 2s_{f_c}^j - 4s_{m_c}^j + s_{f_c}^j s_{m_c}^j + s_c^j s_{f_c}^j - s_c^j s_{m_c}^j + (2s_{m_c}^j - 2s_{f_c}^j)\prod_{r_i \in R(c_j)}\frac{1+s_r^i}{2}) \tag{39}$$

The Hamiltonian of the problem can be simply obtained according to deduced Ising model, by replacing the spin variable quantity to the quantum operator. In this paper, the Pauli-Z operator is utilized to construct the cost Hamiltonian. By summing the above-mentioned terms, the cost Hamiltonian of fault diagnosis problem considering the failed/mal operation of PR and CB, as well as the contradictory logic, is constructed.

**Experiments settings.** In this paper, case studies are conducted on the test system used by [38], as shown in Fig. 2. The system consists of 28 components, 40 CBs and 84 PRs:
- 28 components: A1, …, A4, B1, …, B8, T1, …, T8, L1, …, L8.
- 40 CBs: CB1, CB2, …, CB40.
- 84 PRs: 36 main PRs (A1m, …, A4m, B1m, …, B8m, T1m, …, T8m, L1m, …, L8m), 48 back-up PRs (T1p, …, T8p, L1Sp, L1Rp, …, L8Sp, L8Rp, L1Ss, L1Rs, …, L8Ss, L8Rs)

where A and B represent the busbar, T represents the transformer, L represents the line, S and R represent the beginning and end of the line respectively (from top to bottom and from left to right), m represents the main protection, p represents the first back-up PR, s represents the second back-up PR. The hyper-parameters of objective functions are $\omega_1 = 1, \omega_2 = 1, \omega_3 = 1, \omega_4 = 40$.

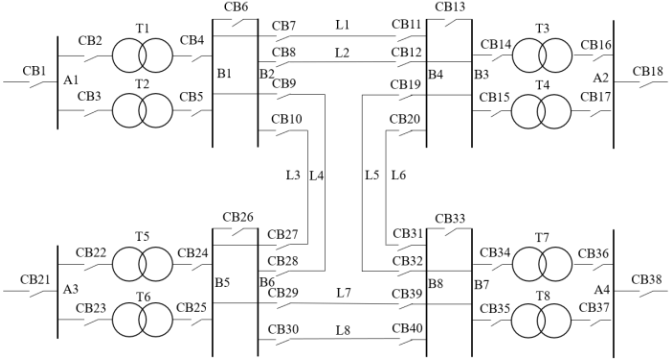

**Fig. 2 The test system which consists of 28 components, 40 CBs and 84 PRs.**

Consider the hardware limits of quantum computer, complex cases cannot be tested using the completed model presented in the Section II. Therefore, this paper also conducted cases based on a simplified model, which do not consider the failed/mal operation information and contradictory logic explicitly [39]. The simplified model's objective function is:

$$\min E(\mathbf{H}) = \sum_{i=1}^{n_r}(r_i \oplus r_i^*) + \sum_{i=1}^{n_c}(c_i \oplus c_i^*) \tag{40}$$

In this way, the number of qubits is equal to the number of suspected components, which greatly reduces the number of qubits required, allowing us to analyze cases with larger number of components.

The quantum simulation environment is PennyLane's "default.qubit" device, which provides a pure state simulation of a qubit-based quantum circuit architecture. The classical solver provided by D-Wave, which can accurately solve the PUBO problem and serve as a benchmark [40].

**Feasibility analysis of proposed method – completed model.** In this test, two fault diagnosis cases are simulated to verify the feasibility of proposed method. The first case considering the failed operation on CB. Suppose a failed operation occurs on a CB. The alarm information is the main PR A1m of busbar A1 operates, CB1 and CB3 trip.

Using topological analysis, the suspected component is the busbar A1, denoted by $D = (d_0)$. The related PR is A1m, denoted by $R = (r_1)$. The related CBs are CB1, CB2 and CB3, denoted by $C = (c_1, c_2, c_3)$. The received alarm information includes $r_1, c_1, c_3$. Therefore, the observed states of PRs and CBs are $R' = (1), C' = (1,0,1)$. According to the small probability event characteristic: $f_{r_1} = 0$, $f_{c_1} = 0$, $m_{c_2} = 0$, $f_{c_3} = 0$. The fault hypothesis $H$ is $(d_0, r_1, c_1, c_2, c_3, f_{c_2}, m_{r_1}, m_{c_1}, m_{c_3})$. The level of QAOA is set to be 10 and the number of iterations is 100.

The test result of proposed method is shown in Fig. 3. The abscissa is the probability of occurrence of each fault hypothesis, and the ordinate is the bit string of fault hypothesis. The optimal fault hypothesis $H^*$ is $(1, 1, 1, 0, 1, 1, 0, 0, 0)$, which means the busbar A1 is faulty, the main PR A1m operates, CB1 and CB3 are tripped, CB2 is failed to trip. The output result is reasonable and consistent with the classical solver.

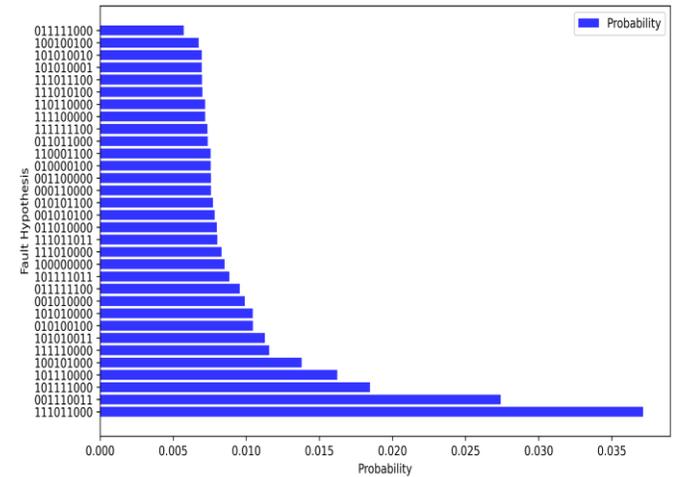

**Fig. 3 Top 32 probable fault hypotheses of Case 1 using proposed method.**

The second case considering the failed operation on PR. Suppose a failed operation occurs on a PR. The alarm information is the first back-up PR of transformer T1 operate, CB2 and CB4 trip.

Using topological analysis, the suspected component is the transformer T1, denoted by $D = (d_0)$. The related PR is T1m and T1p denoted by $R = (r_1, r_2)$. The related CBs are CB2 and CB4, denoted by $C = (c_1, c_2)$. The received alarm information includes $r_2, c_1, c_2$. Therefore, the observed states of PRs and CBs are $R' = (0,1), C' = (1,1)$. According to the small probability event characteristic: $m_{r_1} = 0$, $f_{r_2} = 0$, $f_{c_1} = 0$, $f_{c_2} = 0$. The fault hypothesis $H$ is $(d_0, r_1, r_2, c_1, c_2, f_{r_1}, m_{r_2}, m_{c_1}, m_{c_2})$. The level of QAOA is set to be 10 and the number of iterations is 130.

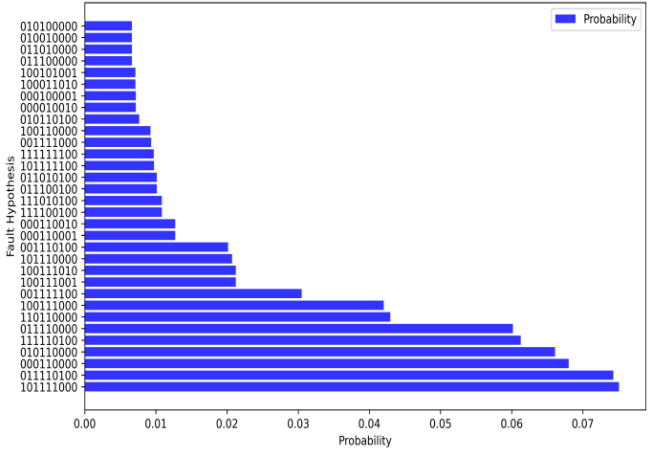

Fig. 4 Top 32 probable fault hypotheses of Case 2 using proposed method.

The test result of proposed method is shown in Fig. 4. The abscissa is the probability of occurrence of each fault hypothesis, and the ordinate is the bit string of fault hypothesis. The optimal fault hypothesis $H^*$ is $(1, 0, 1, 1, 1, 1, 0, 0, 0)$, which means the transformer T1 is faulty, the main PR T1m is failed to operate, the first back-up PR T1p operates. CB2 and CB4 are tripped. The output result is also reasonable and consistent with the classical solver.

**Analysis of small probability event characteristic.** To verify that the SPEC method can help to reduce the number of quantum calculations required while ensuring the accuracy, comparison between the proposed model with and without SPEC is conducted under case 1. The result of the model with SPEC is shown in the previous part. When SPEC is not used, the fault hypothesis $H$ becomes: $(d_0, r_1, c_1, c_2, c_3, f_{r_1}, f_{c_1}, f_{c_2}, f_{c_3}, m_{r_1}, m_{c_1}, m_{c_2}, m_{c_3})$, which requires 13 qubits. However, in our proposed model with SPEC, only 9 qubits are required. In addition, as shown in TABLE I, the circuit depth and number of quantum gates of the model with SPEC are 59 and 83, while that of the model without SPEC are 163 and 198, which is much larger. Next, the result of the model without SPEC is obtained through our proposed framework. The level of QAOA is set to be 10 and the number of iterations is 100.

TABLE I
USING SPEC VS. NOT USING SPEC FOR CASE 1

| METHOD | QUBITS | CIRCUIT DEPTH | QUANTUM GATES | SIMULATED TIME (S) |
|---|---|---|---|---|
| SPEC | 9 | 59 | 83 | 51.47 |
| WITHOUT SPEC | 13 | 163 | 198 | 182.48 |

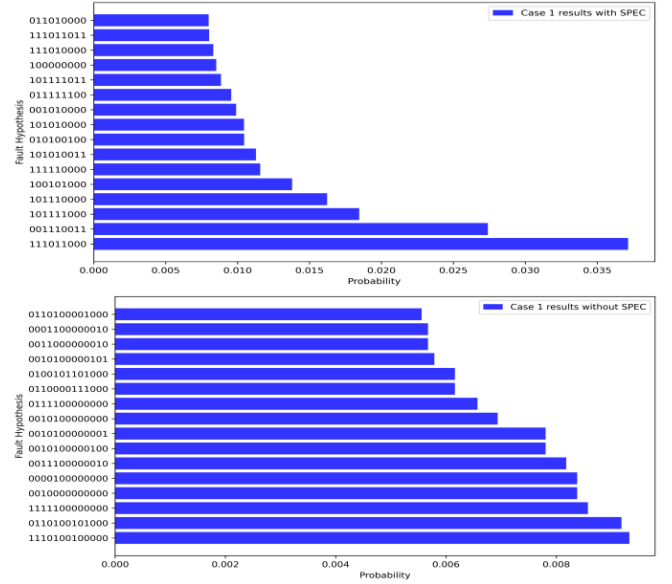

Fig. 5 Top 16 probable fault hypotheses of Case 1 with SPEC and without SPEC.

The result of the model without SPEC is presented in Fig. 5. The result is also reasonable, however, the probability of the optimal solution obtained by our proposed model is about 4 times that of the model without SPEC. In addition, the probability of this optimal solution in our model is obviously greater than other fault hypotheses, while for the results without SPEC, there are many interference solutions with similar probabilities. Besides, the simulated time for the model with SPEC and without SPEC is 51.47s and 182.48s respectively, as shown in TABLE I. Our model is obviously faster than the model without SPEC.

**Feasibility analysis in different problem complexities – simplified model.** In order to prove that the proposed framework can obtain accurate fault diagnosis results under problems of different complexity, this paper has conducted 30 tests for different complexities of the test system shown in Fig. 2 using the proposed simplified model. Due to space limitations, only six cases are listed, as shown in Table II.

TABLE II
TEST CASES USING SIMPLIFIED MODEL

| TEST NUMBER | ALARMS | DIAGNOSIS RESULTS |
|---|---|---|
| 1 | A4m, T7p operate; CB34, 36, 37, 38 are tripped | A4, A7 are faulted |
| 2 | B1m, L2Rs, L4Rs operate; CB4, 5, 7, 12, 28 are tripped | B1 is faulted |
| 3 | B2m, B6m, L4Rs operate; CB6, 7, 8, 10, 26, 28, 30 are tripped | B2, B6 are faulted |
| 4 | A3m, B6m, L4Sm, L4Rp, L3Ss operate; CB 9, 10, 21, 22, 23, 28, 30 are tripped | A3, B6, L4 are faulted |
| 5 | A3m, T5p operate; CB21, 22, 23, 24, 25 are tripped | A3, T5 are faulted |
| 6 | B5m, B7m, B8m, L6Ss, L7Ss operate; CB20, 24, 25, 26, 27, 29, 32, 34, 35, 40 are tripped | B5, B7, B8 are faulted |

TABLE III
COMPARISON OF DIFFERENT PROBLEM COMPLEXITIES

| TEST NUMBER | PROBLEM COMPLEXITY | QUBITS | CIRCUIT DEPTH | QUANTUM GATES | SIMULATED TIME (S) |
|---|---|---|---|---|---|
| 1 | Low | 2 | 6 | 9 | 4.28 |
| 3 | Medium | 4 | 20 | 29 | 10.66 |
| 6 | High | 5 | 59 | 71 | 32.63 |

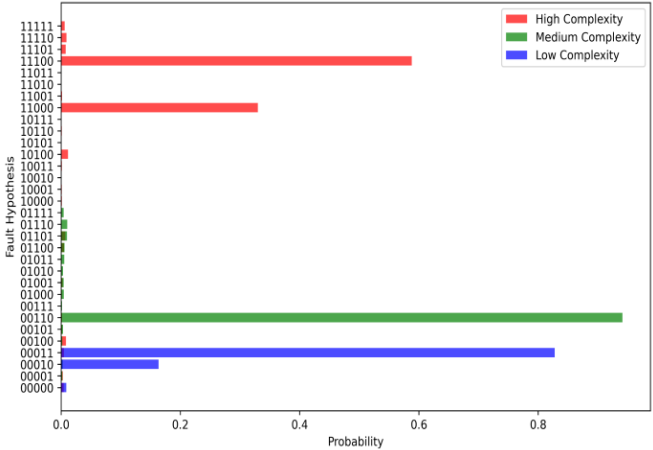

**Fig. 6 The results of test 1 (Low Complexity), test 3 (Medium Complexity) and test 6 (High Complexity).**

We compared three tests (test 1, 3 and 6) of different complexities (Low, Medium, High) in Table II. The number of qubits, circuit depth, quantum gates and simulated time of these three tests are shown in Table III. The results of these tests using the proposed framework are shown in Fig. 6. For all these tests, the optimal solution has high probability and is significantly higher than the other solutions, which shows the feasibility and robustness of our proposed framework under different complexity and the diagnosis results are consistent with the classic solver.

**Comparison between symmetric equivalent decomposition and global variable substitution.** In order to verify that the proposed SED method can reduce the complexity of the quantum circuit and improve the efficiency while ensuring the accuracy of the results, we use the SED method to solve test 3 in TABLE II as an example, and compared the results with that obtained by using an existing method called Global Variable Substitute (GVS) [34], which is also used to solve the PUBO problem with QAOA. From Fig. 7 and Table IV, we find that the number of qubits and quantum gates used in our model is about half of that in the GVS-based model, and the depth of our model is also obviously smaller. In addition, the running time of our model on the simulator is 10.66s, while that of the GVS-based model is 21.71s, which is about two times larger than our method. In fact, the advantage of our method becomes more pronounced as the problem size grows larger, since the quantum gates required by our decomposition-based method are only linear with the number of qubits and do not require any additional auxiliary qubits, whereas the GVS method requires many auxiliary qubits to satisfy the mathematical constraints.

The results of the two models are shown in Fig. 8. The optimal fault diagnosis of SED-based method is '0110', which means components B2 and B6 are faulty, while that of GVS-based method is '01101001', which corresponds to the same diagnosis result. Although both getting the correct solution, our proposed SED-based model has significantly better performance. First, the probability of the correct optimal solution in our model is obviously greater than other invalid solutions. However, in the results of the GVS-based model, there are many interference solutions with similar probabilities, such as '11110000', which is not a reasonable diagnosis result. Besides, the results of the GVS model are affected by more penalized hyperparameters. Therefore, the proposed SED-based method is much more efficient and robust.

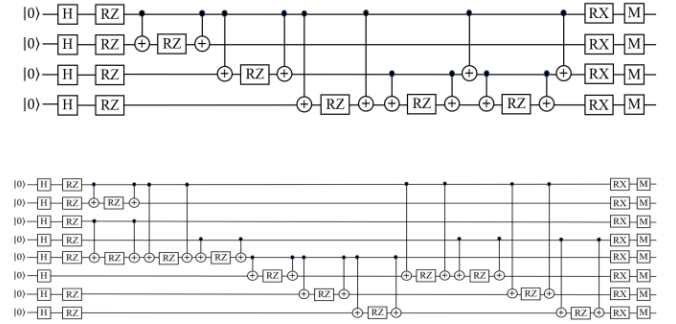

**Fig. 7 The quantum circuit of test 3 using Symmetric Equivalent Decomposition (SED) and Global Variable Substitution Method (GVS).** The upper figure shows the circuit using SED, while the lower figure shows the circuit using GVS.

TABLE IV
COMPARISON BETWEEN SYMMETRIC EQUIVALENT DECOMPOSITION AND GLOBAL VARIABLE SUBSTITUTION

| METHOD | QUBITS | CIRCUIT DEPTH | QUANTUM GATES | SIMULATED TIME (S) |
|---|---|---|---|---|
| SED | 4 | 20 | 29 | 10.66 |
| GVS | 8 | 33 | 56 | 21.71 |

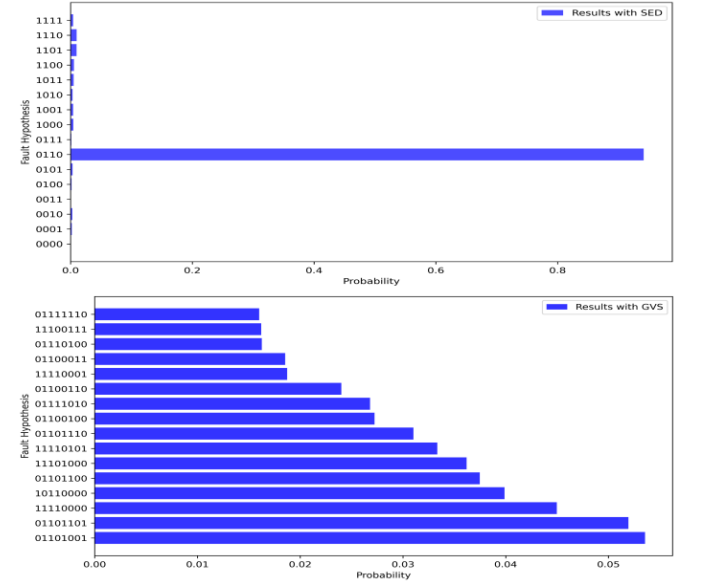

**Fig. 8 The results of test 3 using SED and GVS method.** The upper figure shows the results with SED-based model, while the lower figure shows the results with GVS-based model (only top 16 probable fault hypotheses are presented due to the space limit).

**Computational time analysis.** To analyze the computational time needed for QAOA in the real-world physical devices, this paper follows the time analysis method in [41]. The optimization run using QAOA is composed of hundreds of iterations, and for each iteration, thousands of repetitions of the same quantum circuit are required to have enough statistic to estimate the cost expectation [42]. Specifically, in this test, the total number of repetitions required is given by (100 iterations per run) × (10000 repetitions for statistics). The time used of a single repetition $T_{sr}$ is depending on the complexity of the quantum circuit and considered to be:

$$T_{sr} = T_P + (circuit\ depth) \times T_G + T_M \qquad (41)$$

where $T_P$ is the time to prepare the initial state, $T_G$ is the average duration of a quantum gate, and $T_M$ is the time required to measure the qubits. The time scales used in this paper reflect realistic projections for state-of-the-art devices based on superconducting circuits: $T_P + T_M = 1\ \mu s$, $T_G = 10\ ns$ [42]–[44]. Multiplying $T_{sr}$ by the number of repetitions, the total computational time of QAOA is obtained. For the classical solver, the actual running time in classic processor is recorded.

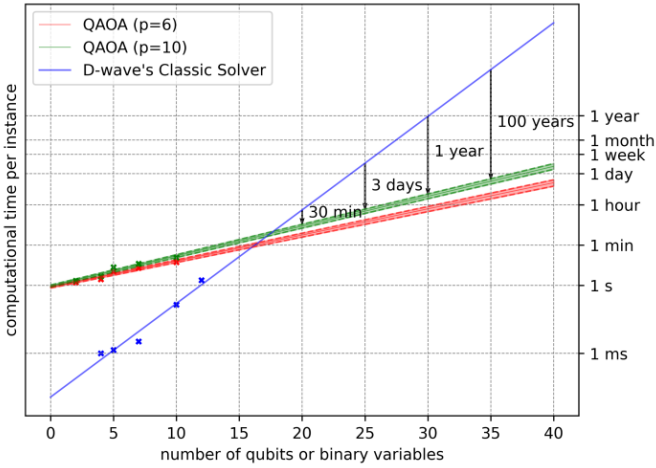

**Fig. 9 Computational time analysis with different amounts of qubits.** Blue markers correspond to the classical baseline (D-Wave's ExactSolver) while red and green marks correspond to the experimental time required by the quantum algorithm QAOA, with p = 6 and p = 10 respectively. The red and green areas are associated with a 95% confidence interval for the prediction of the QAOA computational time per instance.

The time needed to solve a single instance is calculated by averaging the results from 5 instances at each problem size, which corresponds to a single marker in Fig. 9. For QAOA, the considered problem sizes are (2, 4, 5, 7, 10). For classic solver, the considered problem sizes are (4, 5, 7, 10, 12), which is not exactly same as that for QAOA to avoid flat baseline and show the actual trend [41].

Following the method in [41], this paper uses an exponential function to fit the relation between cost time and the number of qubits or binary variables. The inference drawn from this extrapolation should be considered as indicating a qualitative trend, given the uncertainty involved in extrapolating from systems of relatively small sizes. Note that exponential curves, as well as smooth curves in general, appear as straight lines in local regions, making it challenging to rule out alternative functional forms for the extrapolation. Nevertheless, it is widely accepted that quantum computers will not be capable of solving NP-hard problems in polynomial time. From Fig. 9, we can find that when the number of qubits to be used reaches about 20, our algorithm is faster than the classical method based on the 95% confidence interval. However, since the method in [41] makes some assumptions, the result can only be viewed as an estimate.

**Conclusions**

Quantum computing offers a new computing paradigm with the potential to efficiently solve large-scale combinatorial optimization problems. This development could prove to be highly valuable in the field of power systems, especially for large-scale systems. In this paper, we proposed a quantum computing-based power system fault diagnosis framework. The mathematical model is defined with PUBO formulation and then reformulate to the Ising model and Hamiltonian, which can be solved using QAOA. To the best of our knowledge, we are among the first to apply quantum computing to the power system fault diagnosis problem. Furthermore, a gate decomposition method and the small probability event are proposed to reduce the complexity of the proposed QAOA-based solving method. Case studies based on the test system demonstrate the ability of the proposed method to obtain the same results as the classical solver. Moreover, the QAOA-based method is able to solve large-scale fault diagnosis problems at a faster rate. Future work would focus on extending the application of the algorithm to more realistic scenarios, such as reducing the depth of quantum circuits, addressing the noise in quantum circuits, and performing error correction of qubits. Another crucial aspect in future work is the development of more stable parameter initialization methods.

**Methods**

**Symmetric equivalent decomposition of multi-z-rotation gate.** As mentioned in the Introduction, the power system fault diagnosis problem can be formulated as a PUBO problem, which can be easily implemented with multi-qubit gates. However, due to the limitations of current hardware, the implementation of multi-qubit gates is difficult and more error-prone.

To address this issue, this paper proposes a method called symmetric equivalent decomposition (SED), which decomposes an $n$ qubits z-rotation gate into $2(n-1)$ CNOT gates and 1 RZ gate. The detailed structure of SED is shown in Fig. 10. The mathematical formula for $n$ qubits-z-rotation gates is $\prod_n^{\otimes} \sigma_z^i$, which is a tensor product $\otimes$ of $n$ Pauli-Z operators $\sigma_z^i$ acting on qubit $i$. The mathematical formulas of CNOT gate and RZ gate are shown as follows:

$$CNOT = \begin{bmatrix} 1 & 0 & 0 & 0 \\ 0 & 1 & 0 & 0 \\ 0 & 0 & 0 & 1 \\ 0 & 0 & 1 & 0 \end{bmatrix} \qquad (42)$$

$$RZ = \begin{bmatrix} e^{-i\frac{\theta}{2}} & 0 \\ 0 & e^{i\frac{\theta}{2}} \end{bmatrix} \qquad (43)$$

The mathematical induction is used to prove the SED.

For $n = 2$, according to [17], a double-qubit z-rotation gate can be decomposed into 2 CNOT gates and 1 RZ gate.

For $n = k$, we need to prove that the inductive step holds, i.e. a k+1-qubit z-rotation gate can be decomposed into 2 CNOT gates and a k-qubit z-rotation gate, as shown in Fig. 11.

Let the formula of $(k+1)$ qubits z-rotation gate be:

$$\sigma_z^{\otimes(k+1)} = \begin{bmatrix} A & & & \\ & -A & & \\ & & -A & \\ & & & A \end{bmatrix} \quad (44)$$

where: $A = \begin{bmatrix} a_1 & & \\ & \ddots & \\ & & a_{2^{k-1}} \end{bmatrix}$ is a diagonal matrix. The value of $a_1, a_2, \cdots, a_{2^{k-1}}$ are 1 or $-1$.

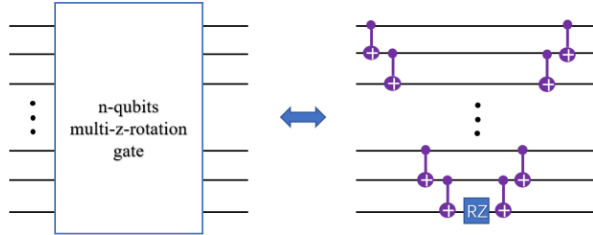

**Fig. 10 The symmetric equivalent decomposition of the multi-z-rotation gate.**

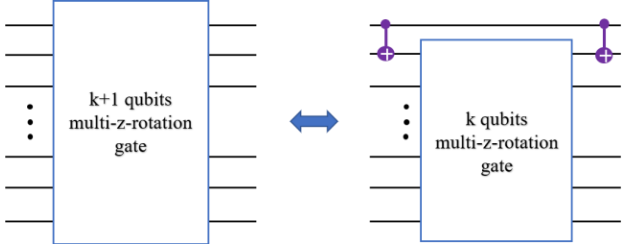

**Fig. 11 Transformation diagram of inductive step.**

The formula of right figure in Fig. 3 is $CNOT \otimes I^{\otimes(k-1)} \otimes I \otimes \sigma_z^{\otimes k} \otimes CNOT \otimes I^{\otimes(k-1)}$

Since

$$CNOT \otimes I^{\otimes(k-1)} = \begin{bmatrix} I_k & 0_k \\ 0_k & 0_{k-1} & I_{k-1} \\ & I_{k-1} & 0_{k-1} \end{bmatrix} \quad (45)$$

$$I \otimes \sigma_z^{\otimes k} = \begin{bmatrix} A & & & \\ & -A & & \\ & & A & \\ & & & -A \end{bmatrix} \quad (46)$$

where $I_k$, $I_{k-1}$ are identity matrix with dimension $2^k$ and $2^{k-1}$, $0_k$, $0_{k-1}$ are zero matrix with dimension $2^k$ and $2^{k-1}$. Therefore, the following equation holds.

$$CNOT \otimes I^{\otimes(k-1)} \otimes I \otimes \sigma_z^{\otimes k} \otimes CNOT \otimes I^{\otimes(k-1)}$$

$$= \begin{bmatrix} I_k & 0_k \\ 0_k & 0_{k-1} & I_{k-1} \\ & I_{k-1} & 0_{k-1} \end{bmatrix} \otimes \begin{bmatrix} A & & & \\ & -A & & \\ & & A & \\ & & & -A \end{bmatrix} \quad (47)$$

$$\otimes \begin{bmatrix} I_k & 0_k \\ 0_k & 0_{k-1} & I_{k-1} \\ & I_{k-1} & 0_{k-1} \end{bmatrix} = \begin{bmatrix} A & & & \\ & -A & & \\ & & -A & \\ & & & A \end{bmatrix} = \sigma_z^{\otimes(k+1)} \quad (48)$$

It means that a k+1-qubit z-rotation gate can be decomposed into 2 CNOT gates and a k-qubit z-rotation gate. Based on mathematical induction, the SED can decompose an $n$ qubits z-rotation gate into $2(n-1)$ CNOT gates and 1 RZ gate.

**Small probability event characteristic of fault diagnosis.**
The proposed objective function in this paper considers the failed/mal operation of PR and CB, which results in a high dimension of the corresponding FH and a large number of qubits in the problem Hamiltonian. For example, a defective scenario comprising of 4 components, 14 PRs, and 8 CBs would require 70 qubits, a capacity beyond the capabilities of the state-of-the-art quantum processor Sycamore developed by Google, which can handle up to 63 qubits. To address this problem, this paper proposes a dimensionality reduction method based on the small probability event characteristics of power system fault diagnosis. By using the proposed method, the mentioned case would only need 48 qubits, which is possible to be handled by the current hardware.

In power system fault diagnosis, small probability events refer to the events that very rarely occur at the same time, and this probability can be normally ignored. For example, for one PR or CB, a mal operation occurs and the alarm information is missed, or the alarm information is falsely reported and the operation is failed. These small probability events fix the relationship of certain variables and can be used to reduce the dimension of the problem. The following two small probability events are applied in this paper:

$$\begin{cases} r_i' = 0, m_{r_i} = 0 \\ r_i' = 1, f_{r_i} = 0 \end{cases} \quad (49)$$

$$\begin{cases} c_i' = 0, m_{c_i} = 0 \\ c_i' = 1, f_{c_i} = 0 \end{cases} \quad (50)$$

As a result, some failed/mal operation variables can be removed from the corresponding FH.

The following is a simple example. Consider the case where the suspected components are $d_0$ and $d_1$, the related PRs are $r_1$ and $r_2$, the related CBs are $c_1, c_2, c_3, c_4$ and $c_5$. The received alarm information includes $r_1, c_1, c_3, c_4$. Therefore, the observed states of PRs and CBs are $R' = (1,0)$ and $C' = (1,0,1,1,0)$. The fault hypothesis $H$ without using small probability event characteristic is ($d_0, d_1, r_1, r_2, c_1, c_2, c_3, c_4, c_5, f_{r_1}, f_{r_2}, f_{c_1}, f_{c_2}, f_{c_3}, f_{c_4}, f_{c_5}, m_{r_1}, m_{r_2}, m_{c_1}, m_{c_2}, m_{c_3}, m_{c_4}, m_{c_5}$), requires 23 qubits. By using small probability event characteristic (SPEC), $f_{r_1} = 0$, $f_{c_1} = 0$, $f_{c_3} = 0$, $f_{c_4} = 0$, $m_{r_2} = 0$, $m_{c_2} = 0$, and $m_{c_5} = 0$, the fault hypothesis $H$ becomes ($d_0, d_1, r_1, r_2, c_1, c_2, c_3, c_4, c_5, f_{r_2}, f_{c_2}, f_{c_5}, m_{r_1}, m_{c_1}, m_{c_3}, m_{c_4}$), which only requires 16 qubits. This paper has proved through experiments that the performance of the processed model will be better.

**Data availability**
The data that support the plots within this article and other findings of this study are available from the corresponding author upon reasonable request.

**Code availability**
Code for this article is available from the corresponding author upon reasonable request.

**Author Contributions**
X.F., H.Z., X.Z., J.Z., T.S., and F.W. conceived and designed the research. X.F. and H.Z. equally contributed to this work in designing the experiments. X.F. derived the core algorithm, designed the framework, wrote the code, conducted the experiments, and wrote the manuscript. H.Z., X.Z., J.Z., T.S., and F.W. revised the manuscript.

**Competing Interests:** The authors declare no competing interests.

**References**

1. V. H. Ferreira *et al.* A survey on intelligent system application to fault diagnosis in electric power system transmission lines. *Electric Power Systems Research* **36**, 135–153 (2016).
2. P. C. Fritzen, J. M. Zauk, G. Cardoso, A. de Lima Oliveira, and O. C. B. de Araújo. Hybrid system based on constructive heuristic and integer programming for the solution of problems of fault section estimation and alarm processing in power systems. *Electric Power Systems Research* **90**, 55–66 (2012).
3. C. Yang, H. Okamoto, A. Yokoyama, and Y. Sekine. Expert system for fault section estimation of power systems using time-sequence information. *International Journal of Electrical Power & Energy Systems* **14**, 225–232 (1992).
4. E. M. Vazquez, O. L. M. Chacon, and H. J. F. Altuve. An on-line expert system for fault section diagnosis in power systems. *IEEE Transactions on Power Systems* **12**, 357–362 (1997).
5. R. N. Mahanty and P. B. Gupta. Application of RBF neural network to fault classification and location in transmission lines. *Generation, Transmission and Distribution, IEE Proceedings* **151**, 201–212 (2004).
6. G. Cardoso, J. G. Rolim, and H. H. Zurn. Application of neural-network modules to electric power system fault section estimation. *IEEE Transactions on Power Delivery* **19**, 1034–1041 (2004).
7. D. Thukaram, H. P. Khincha, and H. P. Vijaynarasimha. Artificial neural network and support vector Machine approach for locating faults in radial distribution systems. *IEEE Transactions on Power Delivery* **20**, 710–721 (2005).
8. M. Barakat, D. Lefebvre, M. Khalil, F. Druaux, and O. Mustapha. Parameter selection algorithm with self adaptive growing neural network classifier for diagnosis issues. *International Journal of Machine Learning and Cybernetics* **4**, 217–233 (2013).
9. J. W. Sheppard and S. G. W. Butcher. A Formal Analysis of Fault Diagnosis with D-matrices. *Journal of Electronic Testing* **23**, 309–322 (2007).
10. A. Perdomo-Ortiz *et al.* Readiness of Quantum Optimization Machines for Industrial Applications. *Phys Rev Appl* **12** (2019).
11. W. Guo, F. Wen, G. Ledwich, Z. Liao, X. He, and J. Liang. An Analytic Model for Fault Diagnosis in Power Systems Considering Malfunctions of Protective Relays and Circuit Breakers. *IEEE Transactions on Power Delivery* **25**, 1393–1401 (2010).
12. Liu, Daobing, Gu, Xueping, Li, and Haipeng. A Complete Analytic Model for Fault Diagnosis of Power Systems. *Proceedings of the CSEE* **31**, 85-92 (2011).
13. S. Wang, D. Zhao, J. Yuan, H. Li, and Y. Gao. Application of NSGA-II Algorithm for fault diagnosis in power system. *Electric Power Systems Research* **175**, 105893 (2019).
14. G. Xiong, D. Shi, J. Zhang, and Y. Zhang. A binary coded brain storm optimization for fault section diagnosis of power systems. *Electric Power Systems Research* **163**, 441–451 (2018).
15. M. Mukherjee. Quantum Computing—An Emerging Computing Paradigm. *Emerging Computing: From Devices to Systems: Looking Beyond Moore and Von Neumann*, 145–167 (2023).
16. M. P. Harrigan *et al.* Quantum approximate optimization of non-planar graph problems on a planar superconducting processor. *Nat Phys* **17**, 332–336 (2021).
17. Y. J. Zhang *et al.* Applying the quantum approximate optimization algorithm to the minimum vertex cover problem. *Appl Soft Comput* **118**, 108554 (2022).
18. Z. Wang, S. Hadfield, Z. Jiang, and E. G. Rieffel. Quantum approximate optimization algorithm for MaxCut: A fermionic view. *Phys Rev A (Coll Park)* **97** (2018).
19. G. G. Guerreschi and A. Y. Matsuura. QAOA for Max-Cut requires hundreds of qubits for quantum speed-up. *Sci Rep* **9**, 6903 (2019).
20. G. E. Crooks. Performance of the Quantum Approximate Optimization Algorithm on the Maximum Cut Problem. arXiv https://doi.org/10.48550/ARXIV.1811.08419 (2018).
21. H. Wang and L.-A. Wu. Ultrafast adiabatic quantum algorithm for the NP-complete exact cover problem. *Sci Rep* **6**, 22307 (2016)
22. A. Bengtsson *et al.* Quantum approximate optimization of the exact-cover problem on a superconducting quantum processor. *arXiv: Quantum Physics* (2019).
23. V. Choi. Different adiabatic quantum optimization algorithms for the NP-complete exact cover problem. *Proceedings of the National Academy of Sciences* **108** (2011).
24. T. Graß. Quantum Annealing with Longitudinal Bias Fields. *Phys Rev Lett* **123** (2019).
25. A. P. Young, S. Knysh, and V. N. Smelyanskiy. Size Dependence of the Minimum Excitation Gap in the Quantum Adiabatic Algorithm. *Phys Rev Lett* **101** (2008).
26. E. Farhi, J. Goldstone, S. Gutmann, J. Lapan, A. Lundgren, and D. Preda. A Quantum Adiabatic Evolution Algorithm Applied to Random Instances of an NP-Complete Problem. *Science* **292**, 472–475 (2001).
27. B. Altshuler, H. Krovi, and J. Roland. Anderson localization makes adiabatic quantum optimization fail. *Proceedings of the National Academy of Sciences* **107**, 12446–12450 (2010).
28. C. Gong, T. Wang, W. He, and H. Qi. A quantum approximate optimization algorithm for solving Hamilton path problem. *The Journal of Supercomputing* **78**, 15381–15403 (2022).
29. N. Nikmehr, P. Zhang, and M. A. Bragin. Quantum Distributed Unit Commitment: An Application in Microgrids. *IEEE Transactions on Power Systems* 37, 3592–3603 (2022).
30. N. Nikmehr, P. Zhang, and M. A. Bragin. Quantum-Enabled Distributed Unit Commitment. *2022 IEEE Power & Energy Society General Meeting (PESGM)*, 1–5 (2022).
31. R. Mahroo and A. Kargarian. Hybrid Quantum-Classical Unit Commitment. *2022 IEEE Texas Power and Energy Conference (TPEC)* (2022).
32. S. Koretsky *et al.* Adapting Quantum Approximation Optimization Algorithm (QAOA) for Unit Commitment. arXiv https://doi.org/10.48550/ARXIV.2110.12624 (2021).
33. A. Perdomo-Ortiz *et al.* Readiness of Quantum Optimization Machines for Industrial Applications. *Phys Rev Appl* **12** (2019).
34. R. Herrman, L. Treffert, J. Ostrowski, P. C. Lotshaw, T. S. Humble, and G. Siopsis. Globally Optimizing QAOA Circuit Depth for Constrained Optimization Problems. *Algorithms* **14** (2021).
35. D. Zhao, X. Zhang, J. Wei, W. Liang, and D. Zhang. Power grid fault diagnosis aiming at reproducing the fault process. *Proceedings of the Chinese Society of Electrical Engineering* **34**, 2116–2123 (2014).
36. W. Guo, F. Wen, Z. Liao, L. Wei, and J. Xin. An Analytic Model-Based Approach for Power System Alarm Processing Employing Temporal Constraint Network. *Power Delivery, IEEE Transactions on* **25**, 2435–2447 (2010).
37. T. Oyama. Fault section estimation in power system using Boltzmann machine. *Proceedings of the Second International Forum on Applications of Neural Networks to Power Systems*, 3–8 (1993).
38. F. Wen and Z. Han. Fault section estimation in power systems using a genetic algorithm. *Electric Power Systems Research* **34**, 165–172 (1995).
39. F. Wen and C.F. Chang. Probabilistic approach for fault-section estimation in power systems based on a refined genetic algorithm. *IEE Proceedings - Generation, Transmission and Distribution* **144**, 160–168 (1997).
40. S. Dutta *et al.* An Ising Hamiltonian solver based on coupled stochastic phase-transition nano-oscillators. *Nat Electron* **4**, 502–512 (2021).
41. G. G. Guerreschi and A. Y. Matsuura. QAOA for Max-Cut requires hundreds of qubits for quantum speed-up. *Sci Rep* **9**, 6903 (2019).
42. A. Kandala *et al.* Hardware-efficient variational quantum eigensolver for small molecules and quantum magnets. *Nature* **549**, 242–246 (2017).
43. R. Barends *et al.* Superconducting quantum circuits at the surface code threshold for fault tolerance. *Nature* **508**, 500–503 (2014).
44. P. J. J. O'Malley *et al.* Scalable Quantum Simulation of Molecular Energies. *Phys Rev X* **6**, 31007 (2016).
45. A. Lucas. Ising formulations of many NP problems. *Front Phys* **2** (2014).
46. P. Vikstål, M. Grönkvist, M. Svensson, M. Andersson, G. Johansson, and G. Ferrini. Applying the Quantum Approximate Optimization Algorithm to the Tail-Assignment Problem. *Phys Rev Appl* **14**, (2020).